\documentclass[preprintnumbers,amsmath,amssymb,nofootinbib]{revtex4}

\usepackage{graphicx}% Include figure files
\usepackage{epsfig}
\usepackage{color}

\newcommand{\beq}{\begin{equation}}
\newcommand{\eeq}{\end{equation}}
\newcommand{\bea}{\begin{eqnarray}}
\newcommand{\eea}{\end{eqnarray}}
\newcommand{\bwd}{\begin{widetext}}
\newcommand{\ewd}{\end{widetext}}

\begin{document}

\title{Mitigation of envelope instability through fast acceleration}
%\title{A comparison of several self-consistent space-charge models for long-term tracking}

%\author{Ji Qiang\thanks{jqiang@lbl.gov}, Lawrence Berkeley National Laboratry, Berkeley, USA }

\author{Ji Qiang}
\email{jqiang@lbl.gov}
\affiliation{Lawrence Berkeley National Laboratory, Berkeley, CA 94720, USA}

\begin{abstract}
The space-charge driven envelope instability can be of great 
danger in high intensity accelerators. Linear accelerators were designed
to avoid this instability by keeping the zero current phase advance per lattice period below $90$ degrees.
In this paper, we studied the acceleration effects 
on the instability in a periodic solenoid and radio-frequency (RF) focusing channel 
and a periodic quadrupole and RF focusing channel using a three-dimensional envelope
model and self-consistent macroparticle simulations. 
Our results show that the envelope instability can be dramatically 
mitigated with a reasonable 
accelerating gradient in both channels. This suggests that the zero current phase advance might be 
above $90$ degrees in linear accelerators where the accelerating gradient is sufficiently high and
opens additional degrees of freedom of transverse and longitudinal focusing parameters
in the accelerator design.
%for stronger transverse and longitudinal focusing

\end{abstract}

\maketitle

\section{Introduction}
In a periodic accelerator system, when the zero current phase advance
through one lattice period is above $90$ degrees, the envelope oscillation of a 
charged particle beam with finite current can become unstable~\cite{ingo83,jurgen}. 
Such an envelope instability driven by the space-charge effects
causes beam size blow up and potential particle losses inside the accelerator. 
In order to avoid the envelope instability, linear accelerators were designed
with zero current phase advance per lattice period below $90$ 
degrees~\cite{apt,stoval,kondo,frank,projx,liz,ess}.
However, the linear accelerator is not a perfect periodic system. In addition
to the machine imperfections such as field amplitude error 
and misalignment error, the use of RF cavity acceleration breaks 
the longitudinal periodicity of the accelerator. The presence of RF acceleration helps
damp the envelope oscillation and also modifies the transverse and longitudinal
focusing strengths and space-charge forces.
Previous studies have shown that resonances in a circular accelerator
can be crossed without significantly affecting the beam quality if the crossing
speed is sufficiently fast~\cite{aiba,lee0,lee1,wang}.
With the use of superconducting RF cavity in the linear accelerator, 
the high accelerating gradient from the cavity may help overcome 
the envelope instability.
%when the zero current phase advance is beyond
%$90$ degrees.
In this paper, we report on mitigating strong envelope instability
of a proton beam through a solenoid-RF lattice and through a quadrupole-RF lattice
with $120$ degree zero current transverse and longitudinal phase advances. 

The envelope instability in a periodic transport channel
has been extensively studied theoretically
 and 
experimentally since 
the 1980s~\cite{ingo83,jurgen,tiefenback,chen,lee2,pakter,fedotov0,fedotov,lund,groening,jeon1,li0,li,ingo2,jeon2,oliver,ingoprab17,ito,yuan,ingo4}. 
%In recent years, there was growing interest in further understanding
%this instability and other structural resonances~\cite{}.
(Some of those studies were summarized in a recently published monograph~\cite{ingobook}.)
It is the lowest order (second-order) collective mode instability driven by the
direct space-charge effects and could present great
danger in high intensity accelerator operation. 
Recently, an analysis of bunched beam stabilities in a periodic
solenoid-RF and in a periodic quadrupole-RF transport channel was done
using a three-dimensional (3D) envelope model without acceleration~\cite{qiangenv}.
%The 3D envelope equations have been used to study the halo particle formation
%mechanism (e.g. particle-core model) for a bunched beam in high intensity 
%accelerators~\cite{bongardt,allen,qiang0,comunian}.
%There, the mismatched envelope oscillation resonates with a test particle and
%drives the particle into large amplitude becoming a halo particle.
%The mismatched envelope oscillation itself is stable in that case. 
%However, when the zero current phase advance per lattice period is
%greater than $90$ degrees, the envelope oscillation itself can become
%unstable.
A number of instability stopbands were identified in that study
when the zero current phase
advance per lattice period is beyond $90$ degrees.
In this paper, 
we study the effects of acceleration on the instability of 
envelope oscillation in these two types of focusing channels.

\section{Three-dimensional envelope instability model with acceleration}

In a linear accelerator, for a 3D uniform density 
ellipsoidal beam,
%inside a periodic focusing channel including acceleration, 
the 3D envelope equations including acceleration are given as~\cite{sacherer,ryne}:
\begin{eqnarray}
	\frac{d^2 X}{d s^2} + (\frac{p_0'}{p_0})\frac{d X}{d s} + k_x^2(s) X - \frac{Ku_0\pi \lambda_3}{l^2}G_{311}(X,Y,u_0T)X -  
	(\frac{\delta}{l p_0})^2 \frac{\epsilon_x^2}{X^3}  =  0   \\
	\frac{d^2 Y}{d s^2} + (\frac{p_0'}{p_0})\frac{d Y}{d s} + k_y^2(s) Y - \frac{Ku_0\pi \lambda_3}{l^2}G_{131}(X,Y,u_0T)Y - 
 (\frac{\delta}{l p_0})^2 \frac{\epsilon_y^2}{Y^3}  =  0    \\
	\frac{d^2 T}{d s^2} + 3(\frac{p_0'}{p_0})\frac{d T}{d s} + k_t^2(s) T - \frac{Ku_0\pi \lambda_3}{l^2}G_{113}(X,Y,u_0T)T - 
 (\frac{\delta}{l p_0 u_0^2})^2 \frac{\epsilon_t^2}{T^3}  =  0   
	\label{3denv}
\end{eqnarray}
with
\begin{eqnarray}
%	I_i(X,Y,Z) = C\int_0^{\infty} \frac{dt}{(e_i^2+t)\sqrt{(X^2+t)(Y^2+t)(\gamma^2 Z^2+t)}}
	G_{lmn}(x,y,z) & = & \frac{3}{2}\int_0^{\infty} \frac{ds}{(x^2+s)^{l/2}(y^2+s)^{m/2}(z^2+s)^{n/2} }
\end{eqnarray}
where $X$ and $Y$ are horizontal and vertical rms beam 
sizes normalized by the scaling length $l = c/\omega$, $T$ is the rms longitudinal
phase ($T=\omega \Delta t$), $\Delta t$ denotes the rms
time of flight difference between the individual particle and the
reference particle, $c$ is the speed of light in vacuum, $\omega$ is the
RF angular frequency, $p_0=mc \gamma_0 \beta_0$ is the reference particle momentum,
the prime denotes derivative with respect to distance $s$, 
$k_x^2$, $k_y^2$, $k_t^2$ represent the external focusing 
forces (($k_x(s)=k_y(s) = qB(s)/(2p_0)$ for solenoids and $k_{x,y}^2(s)= \pm qG(s)/p_0$ for quadrupoles, where $B$ is the solenoid root mean-squared magnetic field along the axis and
$G$ is the quadrupole gradient, $k_t^2 = \omega q E_0 T_{tr} \sin(-\phi_s)/(mc^3\beta^3\gamma^3)$ for longitudinal RF 
focusing~\cite{wrangler}),  $u_0 = \gamma_0 \beta_0$,
$\delta = mc$,
$\epsilon_x$, $\epsilon_y$, and $\epsilon_t$ are
normalized rms emittances,  
and 
$K$ is the generalized perveance associated with the space-charge strength 
given by:
\begin{eqnarray}
	 K & = & \frac{qI}{2 \pi \epsilon_0 p_0 v_0^2 \gamma_0^2}
\end{eqnarray}
where $I$ is the average current of the beam, $q$ is the charge of the particle,
$\epsilon_0$ is the vacuum permittivity, 
$v_0$ is the speed of the reference particle,
and $\gamma_0$ is the relativistic factor of the reference particle.
The quantity $\lambda_3$ is a constant depending on the distribution of the beam.
It was pointed out in reference~\cite{sacherer} that the space-charge
form factor $\lambda_3=1/5\sqrt{5}$ for a uniform distribution depends only weakly
on the type of distributions and is $1.01/5\sqrt{5}$ for a parabolic distribution and $1.05/5\sqrt{5}$ for a Gaussian distribution. 

The above envelope equations can be linearized with respect to matched solutions
as:
\begin{eqnarray}
	X(s) & = & X_0(s) + x(s)  \\
	Y(s) & = & Y_0(s) + y(s)  \\
	T(s) & = & T_0(s) + t(s)  
\end{eqnarray}
where $X_0$, $Y_0$ and $Z_0$ denote the matched envelope solutions 
and $x$, $y$ and $t$ denote small perturbations
\begin{equation}
	x(s) \ll X_0(s), \ \ \ \ y(s) \ll Y_0(s), \ \ \ \ t(s) \ll T_0(s)
\end{equation}

Let $\xi = (x,p_x,y,p_y,t,p_t)^T$, 
the equations of motion for the perturbations are given by:
\begin{eqnarray}
	\frac{d \xi}{d s} & = & A_{6}(s) \xi(s)
	\label{3deq}
\end{eqnarray}
with the matrix $A_6$ given by:
\begin{eqnarray}
	A_{6}(s) & = & \begin{pmatrix}
		0 & \frac{\delta}{lp_0(s)} & 0 & 0 & 0 & 0 \\
		a_{21}(s) & 0 & a_{23}(s) & 0 & a_{25}(s) & 0 \\
		0 & 0 & 0 &  \frac{\delta}{lp_0(s)} & 0 & 0 \\
		a_{23}(s) & 0 & a_{43}(s) & 0 & a_{45}(s) & 0 \\
		0 & 0 & 0 & 0 & 0 & \frac{\delta}{lp_0(s) u_0^2(s)} \\
		a_{25}(s) & 0 & a_{45}(s) & 0 & a_{65}(s) & 0 \\
		 \end{pmatrix}
\end{eqnarray}
where
\begin{eqnarray}
	a_{21}(s) & = & -\frac{l p_0}{\delta} k_x^2 - 3 \frac{\delta}{l p_0} \frac{\epsilon_x^2}{X_0^4} 
	- \frac{Ku_0^2\pi\lambda_3}{l}(3X_0^2G_{511}-G_{311}) \\
a_{23}(s) & = & -\frac{Ku_0^2\pi\lambda_3}{l} X_0 Y_0 G_{331} \\
a_{25}(s) & = & -\frac{Ku_0^4\pi\lambda_3}{l} X_0 T_0 G_{313} \\
	a_{43}(s) & = &  -\frac{l p_0}{\delta}k_y^2 - 3 \frac{\delta}{l p_0} \frac{\epsilon_y^2}{Y_0^4} 
       - \frac{Ku_0^2\pi\lambda_3}{l}(3Y_0^2G_{151}-G_{131})	 \\
a_{45}(s) & = & -\frac{Ku_0^4\pi\lambda_3}{l} Y_0 T_0 G_{133} \\
	a_{65}(s) & = &  -\frac{l p_0u_0^2}{\delta}k_z^2 - 3\frac{\delta}{l p_0 u_0^2} \frac{\epsilon_t}{T_0^4} -  \frac{Ku_0^4\pi\lambda_3}{l}(3u_0^2T_0^2G_{115}-G_{113})
\end{eqnarray}

Let solution $\xi(s) = M_{6}(s) \xi(0)$, substituting this equation into Eq.~\ref{3deq}
results in
\begin{eqnarray}
	\frac{d M_6(s)}{ds} & = & A_6(s) M_6(s)  
\end{eqnarray}
where $M_6(s)$ denotes the $6\times 6$ transfer matrix solution of 
$\xi(s)$ and $M_6(0)$ is a $6\times 6$ unit matrix. 
The above ordinary differential equation can be solved using the matched
envelope solutions and numerical integration.
%Similar to the 2D envelope instability model, 
The stability of these envelope 
perturbations is determined by the eigenvalues of the transfer matrix $M_6(L)$
through one lattice period.
For the envelope oscillation to be stable, 
all six eigenvalues (three pairs) of the $M_6(L)$ have to stay on the unit 
circle.
The amplitude of the eigenvalue gives the envelope mode growth 
(or damping) rate through one lattice period, while the phase of 
the eigenvalue yields the 
mode oscillation frequency. When the amplitude of any eigenvalue is
greater than one, the envelope oscillation becomes unstable.

\section{Mitigation of the envelope Instability in a solenoid and RF channel}
\begin{figure}
%\begin{figure}[htb]
%   \vspace*{-.5\baselineskip}
   \centering
   \includegraphics*[angle=0,width=180pt]{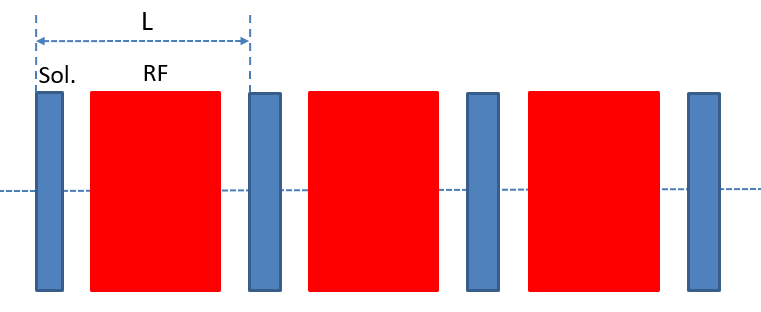}
	\caption{Schematic plot of a periodic solenoid and RF channel. }
   \label{sol3d}
%   \vspace*{-\baselineskip}
\end{figure}
\begin{figure}
   \centering
   \includegraphics*[angle=0,width=180pt]{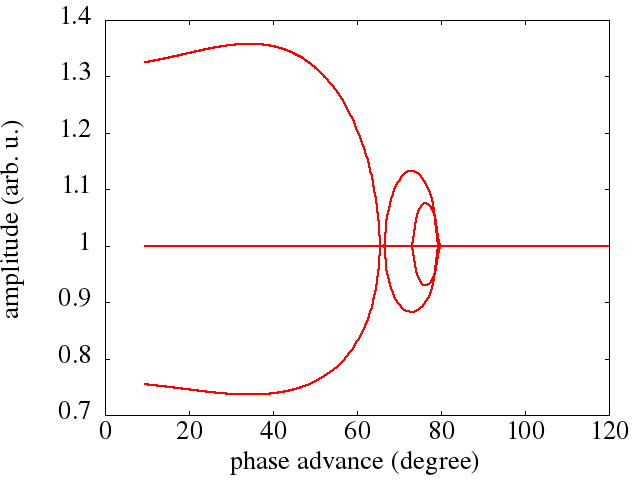}  
   \includegraphics*[angle=0,width=180pt]{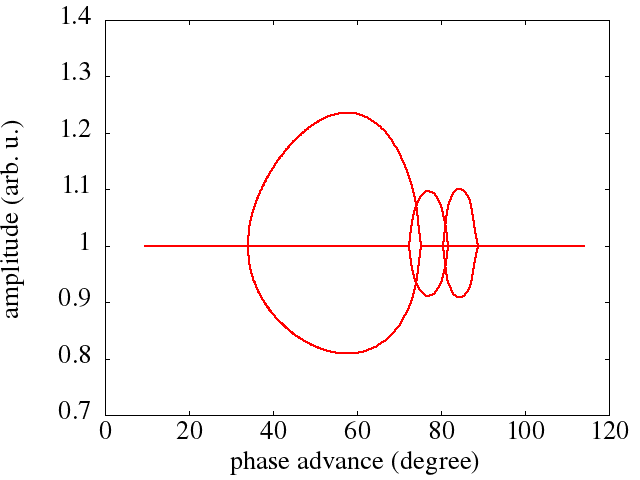}  
   \caption{The 3D envelope mode growth rate amplitudes as a function of 
	depressed transverse phase advance with $120$ 
degree zero current transverse and longitudinal phase advances
	in the periodic solenoid-RF channel without acceleration (left) and
	with $8$ MV/m accelerating gradient (right).
	}
   \label{sol3damp}
\end{figure}

%\begin{figure}
%   \centering
%   \caption{The 3D envelope mode growth rate amplitudes as a function of 
%	depressed transverse phase advance with $120$ 
%degree zero current transverse and longitudinal phase advances
%	in the periodic solenoid-RF channel with $8 MV/m$ accelerating gradient.
%	}
%   \label{sol3damp8}
%\end{figure}
We first studied the mitigation of the envelope instability in a transverse solenoid focusing
and longitudinal RF accelerating and focusing channel.
A schematic plot of this channel is shown in Fig.~\ref{sol3d}.
Each period of the accelerator lattice consists of a $0.1$ meter solenoid, a $0.1$ meter drift,
a $0.4$ meter RF cavity, and $0.1$ meter drift.
%The total length of the period is $0.7$ meters.
The proton bunch has a kinetic energy of $10$ MeV and normalized rms emittances
of $0.2$ um in all three directions.
%$0.2$ um, and $0.2$ um in horizontal, vertical, and longitudinal
%directions respectively.
Figure~\ref{sol3damp} shows the envelope mode growth rate amplitudes 
as a function of depressed transverse phase advance inside the 
above lattice with $120$ degree 
zero current transverse and $120$ degree zero current longitudinal phase advances
without and with $8$ MV/m accelerating gradient inside the RF cavities.
Here, the smaller depressed phase advance corresponds to the stronger space-charge
defocusing effects and the higher proton beam bunch intensity.
Without acceleration, the beam shows strong envelope instability. There are three 
instability stopbands below about $80$ degree depressed transverse phase advances.
The stopband below $65$ degree depressed tune is due to the confluent 
resonance between two envelope oscillation modes.
The other two stopbands are due to the half-integer parametric
resonance between the focusing lattice and the envelope
oscillation modes.
With the presence of acceleration, 
the above envelope instability stopbands
are significantly modified.
The width of stopband is reduced
from about $80$ degrees to about $50$ degrees.
The instability growth rate maximum amplitude is also
reduced from nearly $1.4$ to about $1.2$.
%For the $120$ degree zero current longitudinal phase advances, 
%The two large stopbands are close to each other along the transverse
%depressed phase advance.
%should I mention about which stop band is caused by which????...
%For the $60$ degree zero current longitudinal phase advances, there is about $20$ degree phase advance
%separation.
%The choice such a large zero current phase advance beyond 90 degree is to
%show the effects of acceleration damping.
%The above envelope instability stopbands can be significantly modified by acceleration inside the
%RF cavities. 
%the acceleration also
%breaks the longitudinal periodicity of the original lattice. 
%The acceleration inside the RF cavity provides damping to the envelope
%oscillation as seen in the envelope equations. In addition, 
%both the transverse and the longitudinal
%focusing strengths depend on the energy of the beam. The space-charge 
%forces also depend on the beam energy. When the beam energy increases due to acceleration,
%the focusing and the space-charge effects become weaker and helps
%the beam move across the original instability stopband.

\begin{figure}[htb]
   \centering
   \includegraphics*[angle=0,width=180pt]{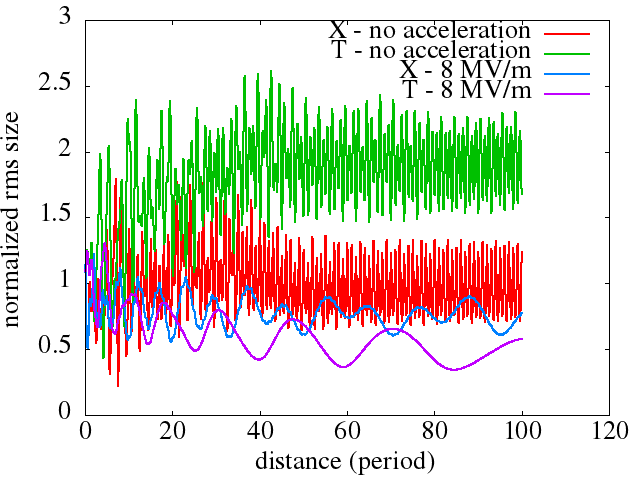} 
   \includegraphics*[angle=0,width=180pt]{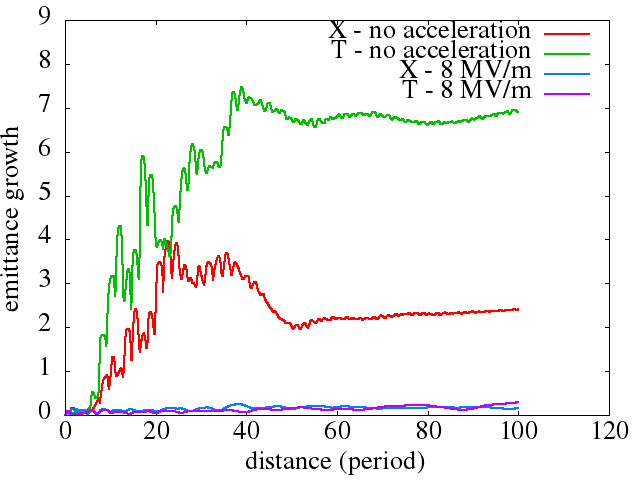} 
	\caption{The horizontal and longitudinal normalized rms size evolution (left) 
	and emittance growth evolution (right) inside 
	the solenoid-RF channel without acceleration and with
	acceleration. The depressed transverse
	phase advance is $40$ degrees with $120$ degree zero current transverse and longitudinal
	phase advances. 
	}
   \label{solrms}
\end{figure}

In order to study the effects of acceleration on the envelope instability, we also carried out 
self-consistent macroparticle simulations using a 3D particle-in-cell code, IMPACT~\cite{impact,impact2}.
We first selected an unstable point inside the depressed phase advance stopband with large 
instability growth rate.
The transverse depressed phase advance for this point is $40$ degrees, which 
represents a strong space-charge tune depression ratio of $0.33$.
%results %in large instability growth rate. 
We assumed $10\%$ mismatch in all three directions of an initial Waterbag distribution
and used about $625,000$ macroparticles and $64\times 64 \times 64$ grid points
in the simulations.
Figure~\ref{solrms} shows the horizontal and longitudinal rms size 
(normalized by the matched rms size) evolution and the emittance growth 
evolution
through the above lattice without and with $8$ MV/m accelerating gradient in the RF cavities.
Here, we assumed that proton beam energy increases linearly inside the RF cavities due to the RF acceleration.
Without acceleration, the transverse and longitudinal rms sizes 
grow quickly up to about $40$ periods due to the instability. 
%becomes unstable.
With the acceleration,
%$8$ MV/m accelerating gradient inside the RF cavities, 
the mismatched
rms envelope oscillations growth is significantly damped and the beam becomes stable. 
Without acceleration, the envelope instability also causes large 
(more than a factor of $3$ and $7$) emittance growth. 
With the acceleration, the emittance growth is small and 
below $30\%$ through this lattice.  

\begin{figure}[htb]
   \centering
   \includegraphics*[angle=0,width=200pt]{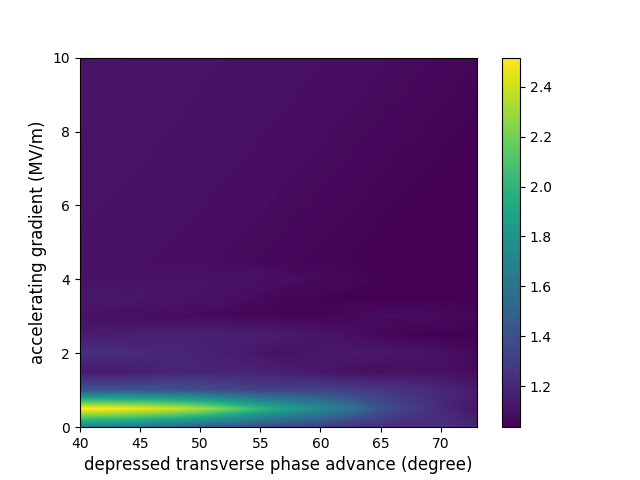} 
   \includegraphics*[angle=0,width=200pt]{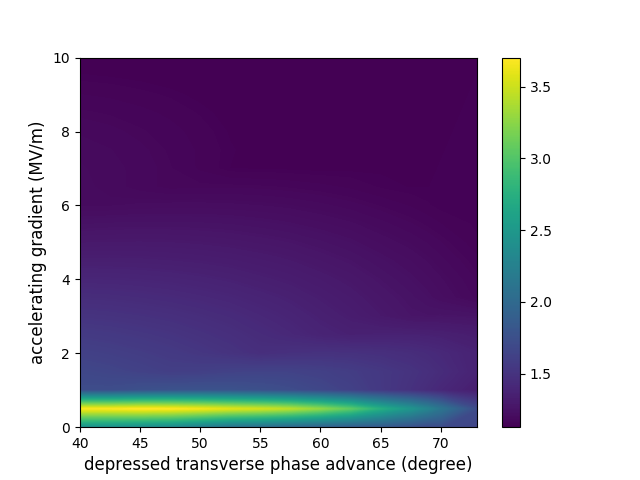} 
\caption{The maximum normalized horizontal rms amplitude (left) and longitudinal
rms amplitude (right) within $100$ lattice periods
as a function of accelerating gradient and depressed transverse
	phase advance in the solenoid-RF channel.}
   \label{sol3dphase}
%   \vspace*{-\baselineskip}
\end{figure}
\begin{figure}[htb]
   \centering
   \includegraphics*[angle=0,width=200pt]{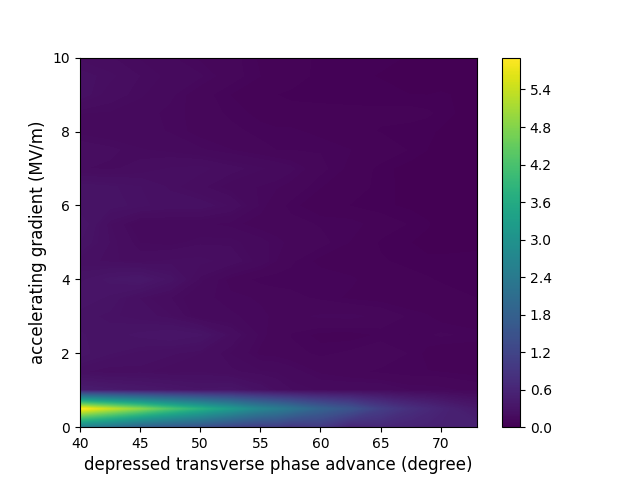} 
   \includegraphics*[angle=0,width=200pt]{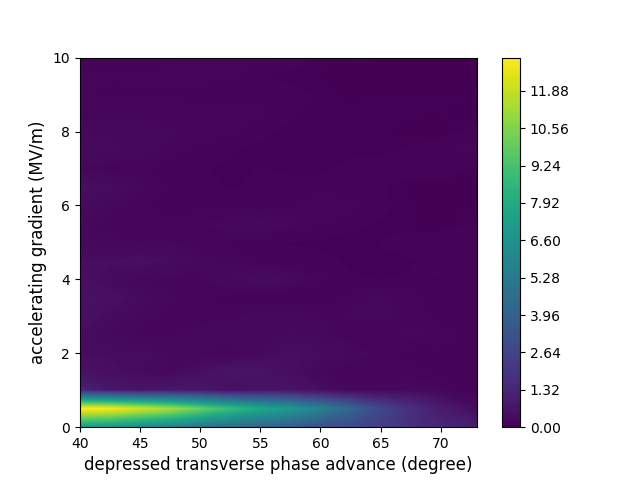} 
\caption{The final horizontal emittance growth (left) and longitudinal
emittance growth (right) at the end of $100$ lattice periods
as a function of accelerating gradient and depressed transverse
	phase advance in the solenoid-RF channel.}
   \label{sol3dphase2}
%   \vspace*{-\baselineskip}
\end{figure}
To see the effects from acceleration systematically, we calculated the maximum
horizontal and longitudinal rms envelope amplitudes
(normalized by the corresponding initial matched rms sizes) 
within $100$ lattice periods and the final emittance growth as a function of
accelerating gradient and depressed transverse 
phase advance in Fig.~\ref{sol3dphase} and Fig.~\ref{sol3dphase2}. Without acceleration
or the accelerating gradient is small, 
the maximum envelope amplitudes 
are significantly greater
than the initial matched rms sizes due
to the envelope instability. Most large amplitudes are located around the left corner
of the above plot
with $40$ degree depressed transverse phase advances and small
accelerating gradient.
This is consistent with the large envelope mode growth rate in the above envelope instability stopband around $40$ degree phase advances. 
With the increase of 
accelerating gradient, the normalized maximum amplitudes decrease and approach  
the initial rms sizes. The mitigation of the instability is also seen in the 
final emittance growth. Most emittance growth occur with accelerating gradient below $1$ MV/m due to the instability. With the increase of accelerating
gradient, the final emittance growth become smaller and approach a few percent level.

The acceleration mitigates the envelope instability 
across the stopbands of depressed phase advance.
The acceleration inside the RF cavity provides damping to the envelope
oscillation as seen in the envelope equations~\ref{3denv}. In addition, 
both the transverse and the longitudinal
focusing strengths depend on the energy of the beam. The space-charge 
forces also depend on the beam energy. When the beam energy increases due to acceleration,
the focusing and the space-charge effects become weaker and helps
move the beam out the original instability stopbands.

\section{Mitigation of the envelope Instability in a quadrupole and RF channel}
Next, we studied the effects of acceleration on the envelope instability
in a transverse quadrupole focusing and longitudinal RF focusing
channel using the same proton beam and computational settings. 
\begin{figure}[htb]
   \centering
   \includegraphics*[angle=0,width=180pt]{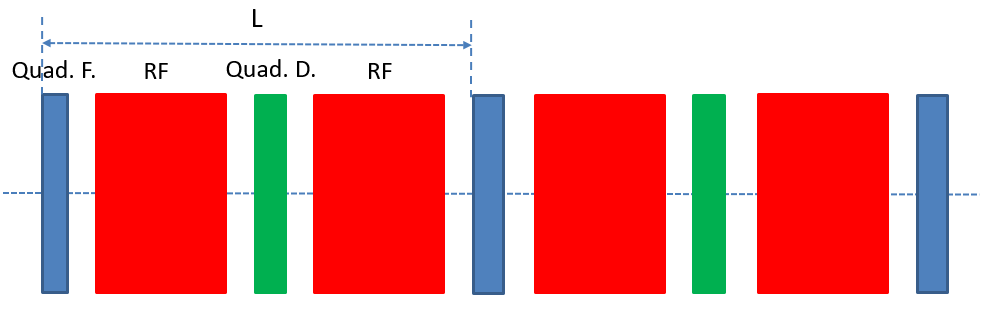}
	\caption{Schematic plot of a periodic quadrupole and RF channel. }
   \label{fd3d}
\end{figure}
\begin{figure}[htb]
   \centering
   \includegraphics*[angle=0,width=180pt]{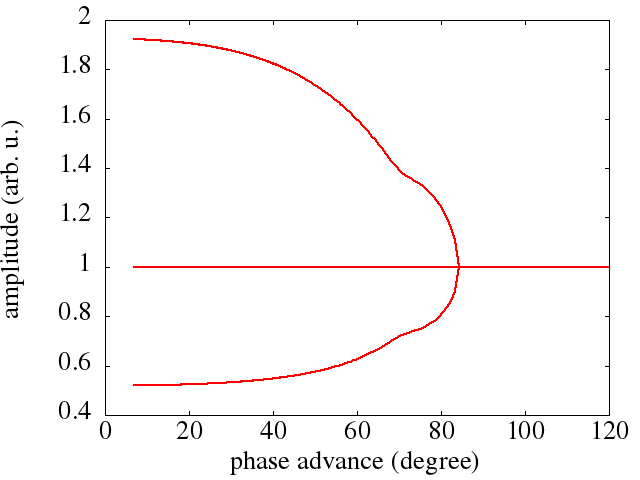} 
   \includegraphics*[angle=0,width=180pt]{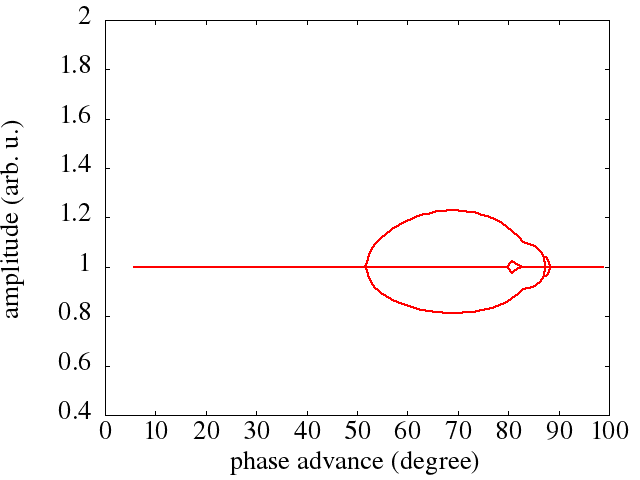} 
   \caption{The 3D envelope mode growth rate amplitudes as a function of 
	depressed transverse phase advance with $120$ 
	zero current transverse and longitudinal phase advances
	in a periodic quadrupole-RF channel without acceleration (left) and
	with $8$ MV/m accelerating gradient (right).
	}
   \label{fd3damp}
\end{figure}
%\begin{figure}[htb]
%   \centering
%   \caption{The 3D envelope mode growth rate amplitudes as a function of 
%	depressed transverse phase advance with $120$ 
%	zero current transverse and longitudinal phase advances
%	in a periodic quadrupole-RF channel with $8 MV/m$ accelerating gradient.
%	}
%   \label{fd3damp8}
%\end{figure}
A schematic plot of this periodic channel is shown in Fig.~\ref{fd3d}.
Each period of the lattice consists of a $0.1$ meter focusing quadrupole,
	a $0.4$ meter RF cavity, a $0.1$ meter defocusing
	quadrupole and another $0.4$
meter RF cavity.
The total length of the period is $1.4$ meters.
Figure~\ref{fd3damp} shows the envelope mode growth rate amplitudes 
as a function of transverse depressed phase advance inside the above lattice with
$120$ degree
zero current transverse and longitudinal phase advances without and with $8$ MV/m
accelerating gradient inside the RF cavities.
Without acceleration, the beam shows strong envelope instability.
Below about $80$ degree phase advances, the
envelope mode growth rate amplitude is greater than one and becomes larger
as the depressed phase advance decreases with stronger space-charge effects. 
%Here, the smaller depressed
%phase advance corresponds to higher bunch intensity inside the beam.
%the beam will
%be unstable. 
This instability stopband is caused by the confluent resonance of
two envelope oscillation modes.
With the presence of acceleration, the envelope instability stopband shrinks 
significantly. Both the stopband width and the growth rate amplitude become 
much smaller in comparison to those without acceleration.

%The above envelope instability stopband is also modified with the presence of acceleration
%in the RF cavities.
We also carried out self-consistent macroparticle simulations 
using the IMPACT code for this lattice.
We first selected a point inside
the instability stopband with $40$ degree depressed transverse phase advances.
\begin{figure}[htb]
   \centering
   \includegraphics*[angle=0,width=180pt]{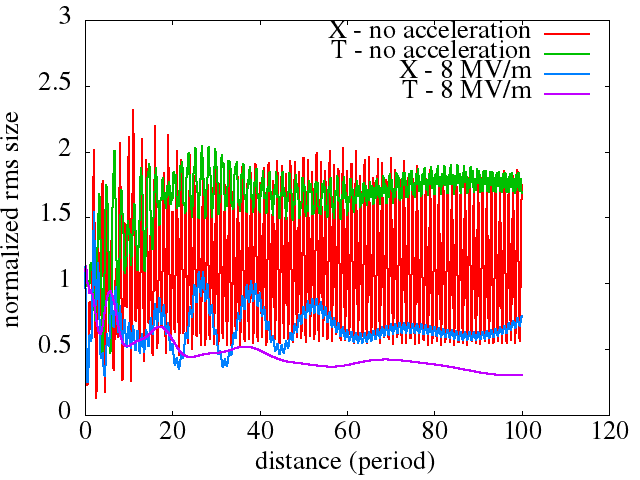} 
   \includegraphics*[angle=0,width=180pt]{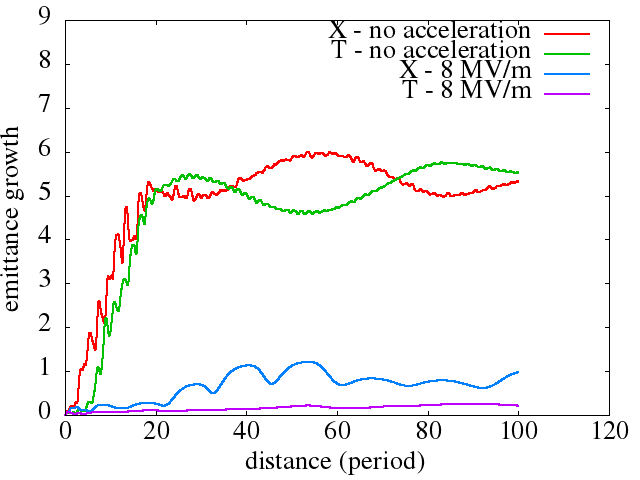} 
   	\caption{The horizontal and longitudinal normalized rms size evolution (left) 
and emittance growth evolution (right) inside 
the quadrupole-RF channel without acceleration and with
acceleration. The depressed transverse
phase advance is $40$ degrees with $120$ degree zero current transverse and longitudinal
phase advances. 
	}
   \label{quadrms}
\end{figure}
Figure~\ref{quadrms} shows the horizontal and longitudinal normalized rms size 
%(normalized by the corresponding matched rms size) 
evolution and emittance growth evolution through the quadrupole-RF channel without and with
$8$ MV/m accelerating gradient inside the RF cavities.
Without acceleration, both horizontal and longitudinal envelope
oscillations are unstable and grow up to about $20$ periods before reaching
saturation.
With the acceleration, except the initial small growth in horizontal plane, 
the rms sizes decrease and become stable through this lattice.
The damping of the longitudinal envelope instability due to
acceleration is faster than that of the transverse envelope instability.
This could be due to a factor of three larger damping rate in the longitudinal 
envelope equation in comparison to those in the transverse envelope equations.
Without acceleration, the normalized emittances increase by more than a 
factor of $5$ due to the envelope
instability. With the acceleration, 
the longitudinal emittance growth is below $30\%$ and
the horizontal growth is below $100\%$. 

Next, we scanned the accelerating
gradient inside the RF cavities with depressed transverse phase advance
inside the above envelope instability 
stopband to systematically study the effects of acceleration. 
\begin{figure}[htb]
   \centering
   \includegraphics*[angle=0,width=200pt]{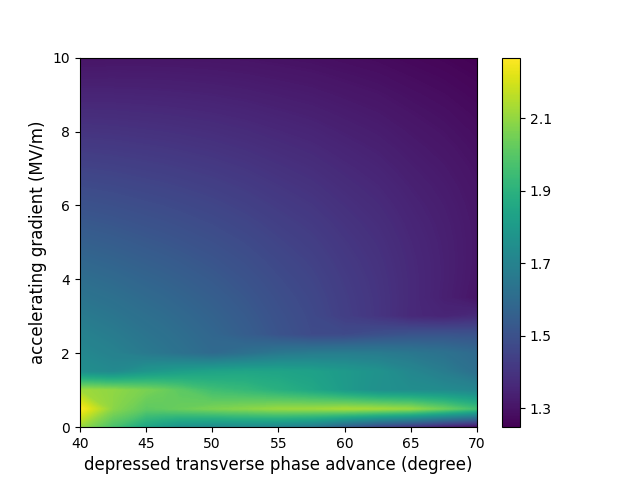} 
   \includegraphics*[angle=0,width=200pt]{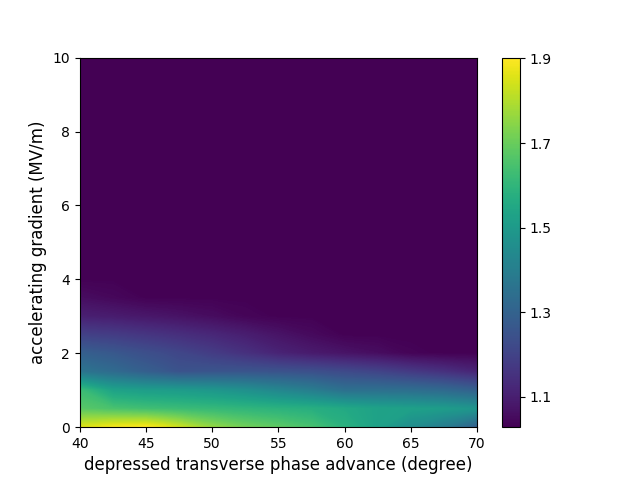} 
\caption{The maximum normalized horizontal rms amplitude (left) and longitudinal
rms amplitude (right) within $100$ lattice periods
as a function of accelerating gradient and depressed transverse
	phase advance in the quadrupole-RF channel.}
   \label{fd3dphase}
\end{figure}
\begin{figure}[htb]
   \centering
   \includegraphics*[angle=0,width=200pt]{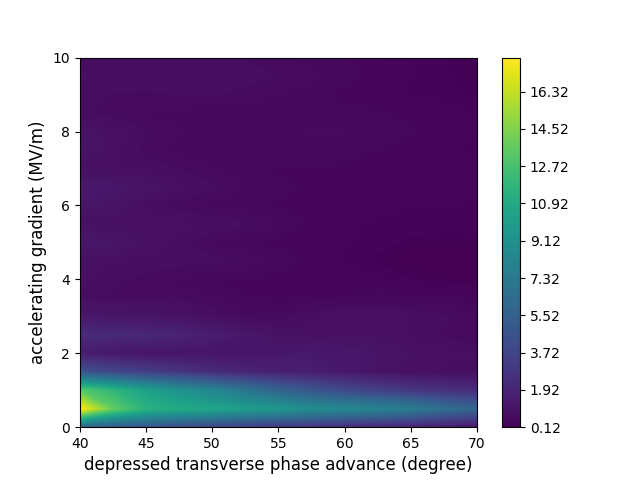} 
   \includegraphics*[angle=0,width=200pt]{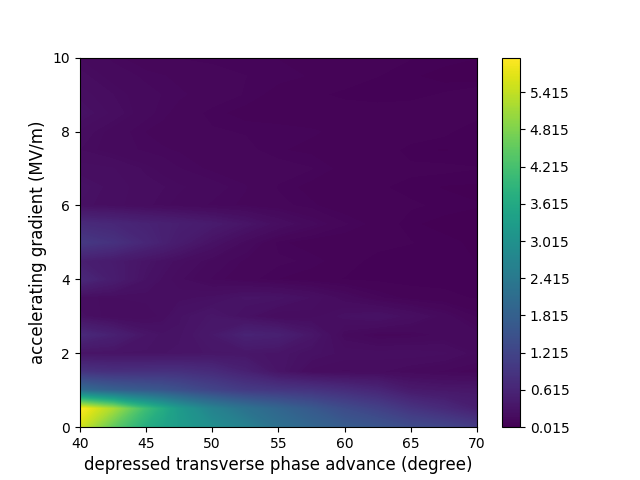} 
\caption{The final horizontal emittance growth (left) and longitudinal
emittance growth (right) at the end of $100$ lattice periods
as a function of accelerating gradient and depressed transverse
	phase advance in the quadrupole-RF channel.}
   \label{fd3dphase2}
\end{figure}
Figures~\ref{fd3dphase} and \ref{fd3dphase2} shows the maximum horizontal and longitudinal
rms amplitudes (normalized by the matched rms sizes) within $100$ lattice
periods and the final emittance growth as a function of
accelerating gradient and depressed transverse phase advance. 
With small or zero accelerating gradient,
the maximum amplitudes are significantly greater than the matched
envelope sizes due to the envelope instability. With the increase of accelerating
gradient, the maximum amplitudes become smaller and approach the matched envelope
sizes. Most large maximum rms amplitudes are located around the left corner
of the plot with small depressed transverse phase advance and accelerating
gradient. The mitigation of the envelope instability is also seen 
in the final emittance growth plot. Most emittance growth occur with
an accelerating gradient below $2$ MV/m. With the increase of the accelerating
gradient, the final horizontal and longitudinal emittance growths become smaller and approach the level of $10\%$.
The acceleration helps mitigate the envelope instability across the stopband
of depressed phase advance in this lattice too.
%and the amplitudes decrease towards larger
%phase advances and high accelerating gradient.

%In both lattices, the damping of the longitudinal envelope instability due to
%acceleration is faster than that of the transverse envelope instability.
%This is probably due to a factor of three damping rate in the longitudinal 
%envelope equation in comparison to those in the transverse envelope equations.
%, which is consistent with the large
%growth rate in the instability stopband. 
%With the accelerating gradient above $4 MV/m$,
%the maximum amplitude become smaller and approach to the matched one across the stopband.

\section{Conclusions}

In this paper, we studied the effects of acceleration on the strong envelope instability using a solenoid-RF lattice and a quadrupole-RF lattice with $120$
degree zero current phase advances in both transverse and longitudinal
directions.
We observed that the beam rms envelope oscillation and final emittance
growth from the envelope instability were dramatically mitigated due to
the RF acceleration in both lattices.
This suggests that the restriction of zero current phase advance per lattice period below
$90$ degrees to avoid the envelope instability in the linear accelerator might be lifted.
This opens additional freedom of choosing transverse and longitudinal focusing 
parameters in the accelerator design
that could result in significant cost savings.
%for stronger transverse and longitudinal focusing

\section*{ACKNOWLEDGEMENTS}
We are grateful for comments by Drs. I. Hofmann and J. Vay and
would like to thank Dr. R. D. Ryne for the 3D envelope code.  
This work was supported by the U.S. Department of Energy under Contract No. DE-AC02-05CH11231 and
used computer resources at the National Energy Research
Scientific Computing Center.

\end{document}